\documentclass[review]{elsarticle}
\usepackage{lineno,hyperref}
\usepackage[english]{babel}
\usepackage[numbers]{natbib}
\usepackage{xcolor}
\usepackage{latexsym,amsmath,amssymb,amsbsy,graphicx,geometry}
\usepackage{epstopdf}
\usepackage{epsfig}
\modulolinenumbers[5]

\begin{document}
	
\begin{frontmatter}
	
\title{A novel view on classification of glass-forming liquids and empirical viscosity model}

\author[adr1,adr2]{Bulat N. Galimzyanov\corref{corauth}}
\cortext[corauth]{Corresponding author}
\ead{bulatgnmail@gmail.com}

\author[adr1,adr2]{Anatolii V. Mokshin}
\ead{anatolii.mokshin@mail.ru} 

\address[adr1]{Kazan Federal University, 420008 Kazan, Russia}
\address[adr2]{Udmurt Federal Research Center of the Ural Branch of the Russian Academy of Sciences, 426067 Izhevsk, Russia}

\begin{abstract}
In the last few decades, theoretical and experimental studies of glass-forming liquids have revealed  presence of universal regularities in the viscosity-temperature data. In the present work, we propose a viscosity model for scaling description of experimental viscosity data. A feature of this model is presence of only two adjustable parameters and high accuracy of experimental data approximation by this model for a wide temperature range. The basis of the scaling description is an original temperature scale. Within this scaling description we obtain the transformed Angell plot, in which the area separating ``fragile'' and ``strong'' glass-formers emerges. The proposed scaling procedure make it possible to reconsider belonging some liquids to the type of ``fragile'' glass-formers. The obtained results form basis for development of a generalized scaling description of crystallization kinetics in supercooled liquids and glasses.
\end{abstract}

\begin{keyword}
	Viscosity \sep Fragility \sep Glass forming ability \sep Amorphous solids \sep Scaled description \sep Universality \MSC[2010] 82B44 \sep 82D20 \sep 82C70
\end{keyword}

\end{frontmatter}

\linenumbers

\section{Introduction}

\begin{figure*}[ht]
	\centering
	\includegraphics[width=15.0cm]{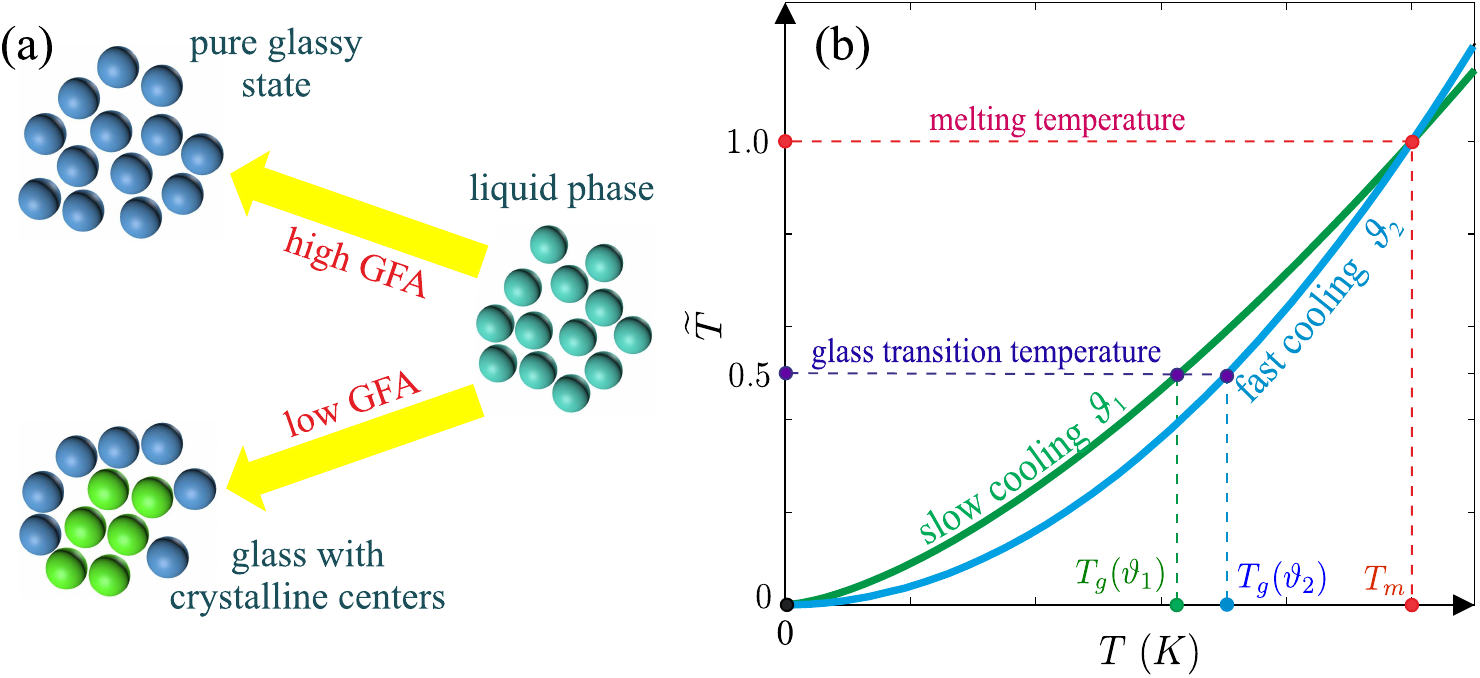}
	\caption{(a) Schematic representation of GFA of glass-forming liquids. (b) Correspondence between reduced temperature $\widetilde{T}$ and absolute temperature $T$ at slow $\vartheta_{1}$ and fast $\vartheta_{2}$ cooling rates. The $\widetilde{T}(T)$-curves are results of Eq.~(\ref{eq_Tscaling}).}\label{fig_1}
\end{figure*}

Amorphous solids have unique physical and mechanical properties that are crucially different from the properties of crystalline analogues~\cite{Avramov_Prado_2003,Wang_Angell_Richert_2006}. The amorphous structure provides high strength, superplasticity, increased corrosion resistance and the improved biocompatibility in the case of some metal alloys~\cite{Binder_Kob_2005,Tantavisut_2018,Brazhkin_2019}. The main physical and mechanical characteristics of amorphous solids are determined by their glass-forming ability (GFA) that is ability to retain the disordered phase without crystallization during rapid cooling of melt [see Fig.~\ref{fig_1}(a)]~\cite{Debenedetti_2001,Lu_Liu_2002,Nascimento_Zanotto_2005,Xu_Liao_2017,Gao_Jian_2019}. For example, silicon dioxide (SiO$_{2}$) and germanium dioxide (GeO$_{2}$) with excellent GFA form a stable homogeneous amorphous structure under normal conditions~\cite{Avramov_2011}. For these materials, it is difficult to obtain crystalline phase rather than glassy one. Bulk metallic glasses have poor GFA, since extremely high cooling rate and rapid heat removal from material are required to suppress the crystallization centers. 

An important key to determine the GFA of glass-forming liquids is correct prediction and understanding the temperature behaviour of viscosity. The structural ordering in supercooled liquids at some temperature $T$, where $T<T_m$, depends on the viscosity $\eta(T)$ or on the structural relaxation time. Typically, the viscosity of liquids increases with decreasing temperature. The loss of molecular mobility with decreasing temperature correlates both with increasing liquid density $\rho$ and with decreasing entropy and thermal activation energy~\cite{Rosenfeld_1980,Lerner_Dyre_2014,Sanditov_Badmaev_2019}. Alba-Simionesco and coworkers have shown that the viscosity of the most glass-forming liquids can be described using the scaled variable $\rho^{\gamma}/T$, where $\gamma$ is the parameter depending on the liquid type~\cite{AlbaSimionesco_2004}. Kelton and coworkers have found for different metallic melts that there is a universal relationship between the viscosity ratio $\log[\eta/\eta_{\infty}]$ ($\eta_{\infty}$ is the extrapolation of viscosity to high temperature limit), and the reduced temperature $T_{A}/T$. The temperature $T_{A}$ corresponds to a state, where cooperative motion of molecules is initiated, and this temperature is strongly correlates with the glass transition temperature $T_{g}$~\cite{Kelton_2015}. Zhang et al. have shown  possibility to evaluate the low-temperature characteristics of liquids based on the high-temperature parameters associated with the activation energy $\Delta E$ and the Arrhenius temperature $T_{A}$~\cite{Jaiswal_Zhang_2016}. On the basis of molecular dynamics simulation results, a quasi-universal relationship for the scaling description of transport coefficients associated with excess entropy was proposed by Dyre and coworkers~\cite{Bell_Dyre_2020}. They have found that the viscosity of model liquid mixtures and binary metal melts as a function of the reduced excess entropy $S_{ex}/k_{B}N$ turns into one universal curve (here $k_{B}$ is the Boltzmann constant, $N$ is the number of atoms or molecules). These findings indicate on presence of common patterns in the temperature dependences of the viscosity of various liquids, and these results can be used to develop a unified viscosity model with the possibility of a scaling description~\cite{Kelton_2015,Sanz_Dyre_2019}. Obviously, such a model should have the minimum number of parameters determined experimentally or from molecular dynamics simulations, and these parameters should be physically argued.

In the present work, we propose an original method for scaling description of experimental viscosity data for the high-density liquids. It is proposed a scaled viscosity model that correctly reproduces the experimental data of completely different types of liquids, including binary and ternary metal melts, silicate, borate, germanium and polymer melts. An obvious advantage of the proposed model is high accuracy of the experimental data approximation using only two adjustable parameters. This is confirmed by comparison of the proposed model with the well-known three-parameter viscosity models. In the proposed model, the reduced temperature scale introduced in Ref.~\cite{Mokshin/Galimzyanov_JCP_2015} is applied. This scale allows one to present uniformly the viscosity-temperature data of liquids with different compositions regardless of the cooling protocol.

\section{Applied methods}

\subsection{Viscosity and kinetic fragility index}

The viscosity $\eta(T)$ of liquids strongly depends on the chemical composition of material, on the heating/cooling protocol, and on the thermodynamic conditions~\cite{Beltyukov_Ladyanov_2019,JCG_2019,Kamaeva_Ryltsev_2020,Galimzyanov_Mokshin_FTT_2020}. For example, the viscosity of low-component silica SiO$_{2}$ and germanium dioxide GeO$_{2}$ increases with decreasing temperature according to the law~\cite{Frenkel_1946}
\begin{equation}\label{eq_fa}
\log\eta(T)=\log\eta_{\infty}+\frac{1}{\ln 10}\frac{\Delta E(T)}{k_{B}T}.
\end{equation}
Here, $\eta_{\infty}$ is the viscosity at infinitely high temperature $T\rightarrow\infty$, $k_B$ is the Boltzmann constant and $\Delta E$ is the activation energy for viscous flow. It turns out that $\Delta E(T)\approx\Delta E_{\infty}$ is the constant at temperatures $T\geq T_{A}$, whereas $\Delta E(T)\propto k_{B}T_{A}(1-T/T_{A})^{8/3}$ performs for a large number glass-forming liquids at $T<T_{A}$~\cite{Tarjus_Viot_2000}. Here, $T_{A}$ is the Arrhenius crossover temperature, at which the transition from Arrhenius to super-Arrhenius $T$-dependence of viscosity is observed. Obviously, for liquids with the strongly pronounced Arrhenius dependence of the viscosity, the high temperature activation energy should be close to the activation energy at $T_{g}$, i.e. $\Delta E_{\infty}\approx\Delta E(T_{g})$~\cite{Jaiswal_Zhang_2016}. For comparison, the $T$-dependence of the viscosity of molecular liquids (for example, salol, o-terphenyl and propylene carbonate) and minerals (such as diopside and anorthite) differs significantly from the Arrhenius behavior: the viscosity changes weakly with temperature near the melting point $T_m$ and the viscosity grows rapidly only near the glass transition temperature $T_g$ and below. Such behavior of the viscosity can be reproduced by the Vogel-Fulcher-Tammann (VFT) empirical equation~\cite{Fulcher_1925,Tammann_Hesse_1926, Nascimento_Aparicio_2007}:
\begin{equation}\label{eq_vft_model}
\log\eta(T)=\log\eta_{\infty}+\frac{1}{\ln10}\frac{B}{T-T_{0}},
\end{equation}
as well as the well-known Avramov-Milchev (AM) equation~\cite{Avramov_Milchev_1988, Avramov_2005}:
\begin{equation}\label{eq_am_model}
\log\eta(T)=\log\eta_{\infty}+\frac{1}{\ln10}\left(\frac{\mathcal{A}}{T}\right)^{\alpha},
\end{equation}
and Mauro-Yue-Ellison-Gupta-Allan (MYEGA) equation~\cite{Mauro_Ellison_2009}:
\begin{equation}\label{eq_myega_model}
\log\eta(T)=\log\eta_{\infty}+\frac{1}{\ln10}\left(\frac{K}{T}\right)\exp\left(\frac{C}{T}\right).
\end{equation}
Here, $T_{0}$ is the ideal glass transition temperature in the VFT equation; $B$, $\mathcal{A}$, $\alpha$, $K$ and $C$ are the fitting parameters. 

The slope of the temperature dependence of the viscosity $\eta(T)$ near the glass transition temperature $T_{g}$ is determined through calculation of the kinetic fragility index $m$~\cite{Kozmidis-Petrovic_2014,Kozmidis-Petrovic_2014_Ceramics,Kelton_Mauro_2014,Martinez_Mauro_2015} 
\begin{equation}\label{eq_m_index}
m=\frac{\partial\log\eta(T)}{\partial(T_{g}/T)}\Big|_{T=T_{g}}.
\end{equation}
Martinez and Angell show that the fragility index $m$ is related with the thermodynamic parameters such as the excess entropy and the specific heat capacity~\cite{Martinez_Angell_2001}. Mauro and coworkers have found a relationship between $m$ and shape of the static structure factor of supercooled liquids~\cite{Kelton_Mauro_2014}. It is established that fragility $m$ can characterize the rate of structural ordering in supercooled liquids near $T_g$. As a rule, the parameter $m$ can take a value from the range [$17$, $200$]~\cite{Angell_1995,Popova_Surovtsev_2011,Novikov_2016,Gutzow_Schmelzer_1995}. The liquids with the fragility index $m$ close to $17$ are usually referred to ``strong'' glass-formers. Such liquids are resistant to structural changes and they have a relatively high viscosity near $T_m$~\cite{Martinez_Angell_2001,Angell_1995,Angell_Leningrad_1989,Angell_Hemmati_2013}. The fragility index is $m=17$ for an ideal ``strong'' glass-former, the viscosity of which follows the law (\ref{eq_fa}) for a wide temperature range: from temperatures of an equilibrium melt to temperatures close to $T_{g}$. Note that glass-formers with $m=17$ are not known till now. For example, the fragility index for SiO$_{2}$ and GeO$_{2}$ is $m\simeq20$ (see Ref.~\cite{Qin_McKenna_2006}) and we have $m\simeq24$ for albite according to Ref.~\cite{Nascimento_Aparicio_2007}. These systems classifies as ``strong''. The liquids with $m$ much higher than $17$ are usually referred to ``fragile''. Such liquids are characterized by extremely low viscosity near $T_{m}$ and spontaneous structural rearrangements near $T_{g}$. This is molecular organic liquids, for example, o-terphenyl with $m\simeq81$ and triphenylphosphate with $m\simeq160$~\cite{Qin_McKenna_2006}. In the case of some polymers and ionic liquids, the fragility index is $m>200$ (for example, polyetherimide $m\approx216$)~\cite{Qin_McKenna_2006,Bohmer_Angell_1994}. It should be noted that Eq.~(\ref{eq_vft_model}) is most often used to calculate the fragility index $m$~\cite{Tammann_Hesse_1926}. Then from (\ref{eq_vft_model}) and (\ref {eq_m_index}) we find
\begin{equation}\label{eq_m_vft}
m^{(VFT)}=\frac{BT_{g}}{(T_g-T_{0})^{2}},
\end{equation}
which allows one to estimate $m$ at known values of the parameters $T_{0}$ and $B$.

\subsection{Unified temperature scale}

Structural ordering in supercooled liquids and glasses occurs at temperatures in the range $0<T<T_m$. This range contains the three critical temperatures: the zeroth temperature $T=0$~K; the glass transition temperature $T_g$ and the melting temperature $T_m$. In our study, the ``critical points'' means such the points on a phase diagram that correspond to phase transitions or to some special (or specific) phase states of the considered system. These ``critical points'' are necessary to modify some range of the phase diagram to a universal form. There are various temperature points, that belong to the considered temperature range $0<T<T_m$, where $T_m$ is the melting temperature. We take three temperature points, where two of them -- the melting temperature $T_m$ and the glass transition temperature $T_g$ -- are independent and usually used to rescale the temperature scale, whilst the zeroth temperature $T=0$~K corresponds to the especial point -- a ground state point -- on a phase diagram. 

The temperatures $T_g$ and $T_m$ are not fixed. These temperatures depend on the type of a liquid and on the liquid cooling protocol~\cite{MG_JETPLett_2019,Mokshin_Galimzyanov_2020}. It is impossible to take into account correctly the universal regularities in the $T$-dependent kinetic characteristics of liquids without fixing the temperatures $T_g$ and $T_m$ in a unified manner for liquids of various types. We have recently proposed a method for scaling the absolute temperature scale in the range from $T=0$~K to $T_{m}$, where the temperatures $T_g$ and $T_m$ are fixed for various liquids~\cite{Mokshin/Galimzyanov_JCP_2015}. This method gives a reduced temperature scale denoted as $\widetilde{T}$~\cite{MG_JETPLett_2019,Mokshin_Galimzyanov_2020}. If the glass transition temperature $T_{g}$ and the melting temperature $T_{m}$ are known for the system, then conversion of the absolute temperature scale to the reduced $\widetilde{T}$-scale is performed according to the following rules (see Fig.~\ref{fig_1}(b)): (i) the melting temperature in $\widetilde{T}$-scale should take the value $\widetilde{T}_m=1$, while the glass transition temperature should take the value $\widetilde{T}_g=0.5$ regardless of the melt cooling protocol; (ii) at the temperature $T=0$~K we have $\widetilde{T}=0$.

The transformation from absolute $T$-scale to the reduced $\widetilde{T}$-scale is carried out using the expression: 
\begin{equation}\label{eq_Tscaling}
\widetilde{T}=K_{1}(T_g,T_m)\left(\frac{T}{T_{g}}\right)+K_{2}(T_g,T_m)\left(\frac{T}{T_{g}}\right)^{2}
\end{equation}
with coefficients
\begin{equation}\label{eq_cond}
K_{1}(T_g,T_m)=\frac{1}{2}-K_{2}(T_g,T_m),\,\,\,
K_{2}(T_g,T_m)=\frac{T_g}{T_m}\left(\frac{T_g}{T_m}-\frac{1}{2}\right)\left(1-\frac{T_g}{T_m}\right)^{-1}.
\end{equation}
For most liquids, the ratio $T_g/T_{m}$ varies from $0.5$ to $0.78$~\cite{Jaiswal_Zhang_2016,Nascimento_Aparicio_2007}. Therefore, the coefficient $K_{2}(T_g,T_m)$ takes positive values from $0$ to $1$: as it follows from Eq.~(\ref{eq_cond}), a larger value of $T_g/T_{m}$ corresponds to a larger value of $K_{2}(T_g,T_m)$~\cite{Mokshin/Galimzyanov_JCP_2015}.

\section{Results}

We have interpreted experimental data on the temperature dependence of the viscosity for $30$ various types of glass-forming liquids~\cite{Wang_Angell_Richert_2006, Nascimento_Aparicio_2007,Ravindren_2014}. The considered liquids differ in composition, type of chemical bonds, and molar mass. Some parameters of these liquids are given in Table~\ref{table_1}.

\begin{figure*}[ht]
	\centering
	\includegraphics[width=15cm]{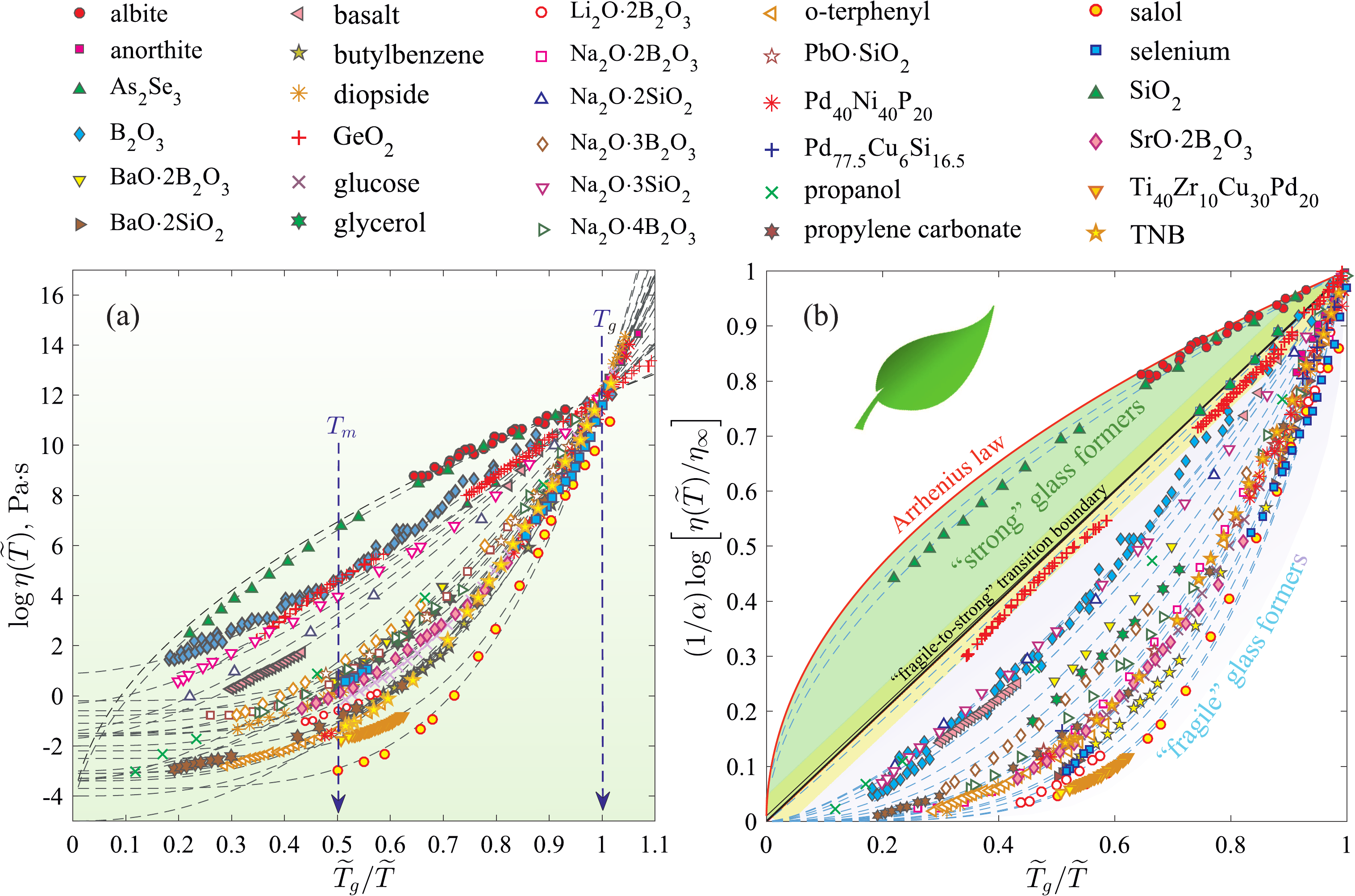}
	\caption{(a) Experimental viscosity data $\log\eta(T)$ of various systems~\cite{Wang_Angell_Richert_2006,Nascimento_Aparicio_2007,Ravindren_2014} as functions of the reduced temperature $\widetilde{T}_{g}/\widetilde{T}$. (a) $(1/\alpha)\log[\eta(\widetilde{T})/\eta_{\infty}]$ vs. $\widetilde{T}_{g}/\widetilde{T}$ plot, where regions of the ``strong'' and ``fragile'' glass-formers are shown. Dashed curves are the result of Eq.~(\ref{eq_1_svm}).}\label{fig_2}
\end{figure*}

Let us present experimental viscosity data on the reduced temperature scale $\widetilde{T}_{g}/\widetilde{T}$ [see Fig.~\ref{fig_2}(a)], where $\widetilde{T}$ is defined by Eq.~(\ref{eq_Tscaling}). As seen in Fig.~\ref{fig_2}(a), the experimental data in the reduced temperature scale differ from the usual $\log\eta(T)$ vs. $T_g/T$ plot. The logarithm of viscosity $\log\eta(\widetilde{T})$ as a function of $\widetilde{T}_{g}/\widetilde{T}$ can be the concave and convex, as well as can be linear with the positive slope. Obviously, to reproduce the experimental data in the $\log\eta(\widetilde{T})$ vs. $\widetilde{T}_{g}/\widetilde{T}$ plot, a power function is needed, where the exponent will regulate the curvature (convexity or concavity) of the viscosity curves. We define such a function as follows: 
\begin{equation}\label{eq_1_svm}
\log\eta(\widetilde{T})=\log\eta_{\infty}+\alpha\left(\frac{\widetilde{T}_{g}}{\widetilde{T}}\right)^{p},
\end{equation}
where $\alpha$ and $p$ are the adjustable parameters. 

Figure~\ref{fig_2}(a) shows that the scaled viscosity (SV) model [see Eq.~(\ref{eq_1_svm}] correctly describes the temperature dependence of the viscosity for all the considered systems. Eq.~(\ref{eq_1_svm}) yields correct asymptotics at $\widetilde{T}\rightarrow\widetilde{T}_{g}$ and $\widetilde{T}\rightarrow\infty$. Indeed, for $\widetilde{T}\rightarrow\infty$ from (\ref{eq_1_svm}) it follows
\begin{equation}\label{eq_alp} 
\log\eta(\widetilde{T}\rightarrow\infty)=\log\eta_{\infty},
\end{equation}
while assuming $\log\eta_{g}\equiv\log\eta(\widetilde{T}=\widetilde{T}_{g})$ for $\widetilde{T}=\widetilde{T}_{g}$ we get
\begin{equation}\label{eq_alpha_1} 
\alpha=\log\eta_{g}-\log\eta_{\infty}.
\end{equation}
Here, $\alpha$ is the so-called lower fragility limit~\cite{Kozmidis-Petrovic_2014, Kozmidis-Petrovic_2014_Ceramics}; parameter $\eta_{g}$ defines the viscosity at the glass transition temperature $T_{g}$ and $\log\eta_{g}=12$ according to definition of $T_{g}$~\cite{Debenedetti_2001,Sanditov_Badmaev_2019}. Avramov shows that the parameter $\alpha$ is related with the activation energy $\Delta E(T_g)$~\cite{Avramov_2005}:
\begin{equation}\label{eq_alpha_2}
\alpha=\frac{1}{\ln 10}\frac{\Delta E(T_g)}{k_{B}T_{g}}.
\end{equation}
Eq.~(\ref{eq_alpha_2}) is valid only for the systems whose temperature dependence of the viscosity is close to the Arrhenius law. We find that for the considered systems given in Table~\ref{table_1} the lower fragility limit takes values in the range $11.1\leq\alpha\leq17.0$. For some ``strong'' inorganic liquids and minerals $\alpha$ is $\approx17$ that is associated with large activation energy for viscous flow. For example, for SiO$_{2}$ we find from Eq.~(\ref{eq_alpha_2}) the activation energy $\Delta E(T_g)\approx471$~kJ/mol at $\alpha=17$ and $T_g=1450$~K. This value is comparable with $\Delta E(T_g)\approx574$~kJ/mol given in Ref.~\cite{Qin_McKenna_2006}. In the case of GeO$_{2}$, the lower fragility limit is $\approx14.0$ due to the relatively low activation energy $\Delta E(T_g)\approx221$~kJ/mol that agrees with the literature value $\Delta E (T_g)\approx258$~kJ/mol~\cite{Qin_McKenna_2006}. Thus, in the case of ``strong'' glass-formers, the parameter $\alpha$ ceases to be adjustable and, according to Ref.~\cite{Jaiswal_Zhang_2016}, this parameter can be determined from Eq.~(\ref{eq_alpha_2}) at the known high-temperature activation energy $\Delta E_{\infty}\approx\Delta E(T_g)$. In the case of the ``fragile'' glass-formers, the parameter $\alpha$ is directly related to the high-temperature activation energy at the Arrhenius transition temperature $T_{A}$, $\alpha\propto\Delta E(T=T_{A})/k_{B}T_{A}$~\cite{Tarjus_Viot_2000}.  

\begin{table}[ht!]
	\begin{center}
		\tiny
		\caption{Properties of considered systems: the glass transition temperature $T_{g}$; the melting temperature $T_{m}$; the ratio $T_{g}/T_{m}$; the parameter $\alpha$ and exponent $p$ of Eq.~(\ref{eq_1_svm}); the fragility index $m$ calculated by Eq.~(\ref{eq_sc_m_index_1}); the experimental values of the fragility index $m_{meas}$ from Refs.~\cite{Wang_Angell_Richert_2006,Angell_Plazek_1993,Senkov_Miracle_2009}; the ratio $\Delta C_{p}(T_g)/\Delta S_{m}$ taken from Ref.~\cite{Wang_Angell_Richert_2006}, where $\Delta C_{p}(T_{g})$ is the heat capacity differences of supercooled liquids relative to the glassy state at $T_g$, $\Delta S_{m}$ is the entropy of fusion. \label{table_1}}
		\begin{tabular}[t]{c|c|ccc|p{30pt}p{34pt}p{34pt}|cc}
			\hline\hline
			Num. & System & $T_{g}$, K & $T_{m}$, K & $T_{g}/T_{m}$ & $\alpha$ & $p$ & $m$ & $m_{meas}$ & $\Delta C_{p}(T_g)/\Delta S_{m}$ \\
			\hline
			1 & albite									& $1087$ & $1393$ & $0.779$ & $17.0\pm0.3$ & $0.51\pm0.03$ & $25.9\pm2.0$ & $27$~\cite{Senkov_Miracle_2009} & -- \\
			2 & anorthite								& $1113$ & $1825$ & $0.609$  & $17.0\pm0.2$ & $2.19\pm0.15$ & $50.0\pm4.1$ & $54$~\cite{Senkov_Miracle_2009} & -- \\
			3 & As$_{2}$Se$_{3}$ 						& $453$  & $633$ & $0.716$   & $12.1\pm0.3$ & $1.54\pm0.11$ & $38.9\pm4.6$ & $40$~\cite{Wang_Angell_Richert_2006} & $1.18$ \\
			4 & B$_{2}$O$_{3}$							& $560$ & $723$ & $0.775$ 	   & $11.1\pm0.1$ & $1.70\pm0.13$ & $54.6\pm3.9$ & $36$~\cite{Wang_Angell_Richert_2006} & $1.21$ \\
			5 & BaO$\cdot2$B$_{2}$O$_{3}$				& $810$ & $1183$ & $0.685$   & $15.3\pm0.2$ & $1.96\pm0.12$ & $54.1\pm4.1$ & -- & -- \\
			6 & BaO$\cdot2$SiO$_{2}$					& $973$ & $1699$ & $0.573$    & $12.2\pm0.2$ & $3.50\pm0.22$ & $51.1\pm4.1$ & -- & -- \\
			7 & basalt									& $988$ & $1473$ & $0.671$  & $13.7\pm0.2$ & $1.60\pm0.11$ & $37.2\pm3.0$ & -- & -- \\
			8 & butylbenzene							& $129$ & $185$ & $0.697$ 	  & $14.2\pm0.1$ & $3.60\pm0.21$ & $97.4\pm6.4$ & -- & -- \\
			9 & diopside								& $995$ & $1664$ & $0.598$   & $13.6\pm0.1$ & $3.15\pm0.19$ & $55.3\pm3.8$ & $59$~\cite{Senkov_Miracle_2009} & -- \\
			10 & GeO$_{2}$								& $810$ & $1378$ & $0.594$    & $14.0\pm0.3$ & $1.17\pm0.05$ & $20.5\pm1.2$ & $20$~\cite{Wang_Angell_Richert_2006} & $0.51$ \\
			11 & glucose								& $295$ & $419$ & $0.704$ 	  & $14.0\pm0.1$ & $3.00\pm0.16$ & $82.7\pm5.0$ & -- & -- \\
			12 & glycerol								& $190$ & $293$ & $0.648$ 	 & $15.0\pm0.2$ & $2.20\pm0.15$ & $50.9\pm4.2$ & $53$~\cite{Wang_Angell_Richert_2006} & $1.44$ \\
			13 & Li$_{2}$O$\cdot2$B$_{2}$O$_{3}$		& $763$ & $1190$ & $0.641$   & $13.5\pm0.1$ & $3.86\pm0.17$ & $78.4\pm4.1$ & -- & -- \\
			14 & Na$_{2}$O$\cdot2$B$_{2}$O$_{3}$		& $748$ & $1015$ & $0.737$   & $13.1\pm0.1$ & $2.90\pm0.12$ & $88.4\pm4.4$ & -- & -- \\
			15 & Na$_{2}$O$\cdot2$SiO$_{2}$				& $728$ & $1146$ & $0.635$   & $13.4\pm0.1$ & $1.52\pm0.08$ & $30.0\pm1.8$ & $45$~\cite{Angell_Plazek_1993} & -- \\
			16 & Na$_{2}$O$\cdot3$B$_{2}$O$_{3}$		& $746$ & $1039$ & $0.718$    & $13.6\pm0.1$ & $2.40\pm0.14$ & $68.9\pm4.6$ & -- & -- \\
			17 & Na$_{2}$O$\cdot3$SiO$_{2}$				& $743$ & $1084$ & $0.685$    & $12.3\pm0.1$ & $1.60\pm0.09$ & $35.5\pm2.3$ & $37$~\cite{Angell_Plazek_1993} & -- \\
			18 & Na$_{2}$O$\cdot4$B$_{2}$O$_{3}$		& $727$ & $1087$ & $0.669$    & $13.5\pm0.2$ & $2.70\pm0.12$ & $61.4\pm3.7$ & -- & -- \\
			19 & o-terphenyl							& $246$ & $329$ & $0.745$   & $15.1\pm0.2$ & $3.10\pm0.18$ & $113.8\pm7.7$ & $81$~\cite{Wang_Angell_Richert_2006} & $2.13$ \\
			20 & PbO$\cdot$SiO$_{2}$					& $673$ & $1037$ & $0.649$  & $12.8\pm0.1$ & $2.76\pm0.13$ & $54.8\pm3.0$ & -- & -- \\
			21 & Pd$_{40}$Ni$_{40}$P$_{20}$				& $583$ & $965$ & $0.604$ 	& $15.3\pm0.3$ & $2.90\pm0.13$ & $58.5\pm3.8$ & -- & -- \\
			22 & Pd$_{77.5}$Cu$_{6}$Si$_{16.5}$			& $645$ & $1058$ & $0.609$  & $16.0\pm0.3$ & $2.70\pm0.12$ & $57.9\pm3.7$ & -- & -- \\
			23 & propanol								& $98$  & $147$ & $0.667$ & $15.4\pm0.3$ & $1.70\pm0.08$ & $43.7\pm2.9$ & $40$~\cite{Wang_Angell_Richert_2006} & $1.25$ \\
			24 & propylene carbonate					& $160$ & $224$ & $0.714$  & $15.1\pm0.2$ & $2.90\pm0.13$ & $90.6\pm5.3$  & $99$~\cite{Wang_Angell_Richert_2006} & $2.12$ \\
			25 & salol									& $215$ & $315$ & $0.683$  & $15.8\pm0.2$ & $3.30\pm0.25$ & $93.1\pm7.5$ & $76$~\cite{Wang_Angell_Richert_2006} & $1.93$ \\
			26 & selenium								& $301$ & $494$ & $0.623$ & $14.1\pm0.2$ & $2.60\pm0.15$ & $49.2\pm3.2$ & $87$~\cite{Wang_Angell_Richert_2006} & $1.16$ \\
			27 & SiO$_{2}$								& $1450$ & $2000$ & $0.725$  & $17.0\pm0.3$ & $0.54\pm0.04$ & $20.0\pm1.9$ & $20$~\cite{Angell_Plazek_1993} & -- \\
			28 & SrO$\cdot2$B$_{2}$O$_{3}$				& $911$ & $1270$ & $0.717$  & $13.6\pm0.2$ & $3.10\pm0.16$ & $88.5\pm5.9$ & -- & -- \\
			29 & Ti$_{40}$Zr$_{10}$Cu$_{30}$Pd$_{20}$	& $687$ & $1280$ & $0.537$ & $14.5\pm0.2$ & $4.40\pm0.27$ & $69.3\pm5.2$ & -- & -- \\
			30 & TNB									& $337$ & $472$ & $0.714$  & $15.4\pm0.3$ & $2.90\pm0.17$ & $92.4\pm7.3$  & $66$~\cite{Angell_Plazek_1993} & -- \\
			\hline\hline
		\end{tabular}
	\end{center}
\end{table}

The proposed SV-model reproduces correctly the shape of the viscosity-temperature curves over a wide temperature range: from temperatures of the equilibrium liquid phase to temperatures near $T_g$. In Eq.~(\ref{eq_1_svm}), the exponent $p$ sets the shape of the viscosity curve and this parameter takes positive values only. So, for $0<p<1$, the result of Eq.~(\ref{eq_1_svm}) is a convex curve, which follows from the condition
\begin{equation}\label{eq_svm_cond_1}
\frac{d^{2}\left(\log\eta(\widetilde{T})\right)}{d\left(\widetilde{T}_{g}/\widetilde{T}\right)^{2}}<0.
\end{equation}
Eq.~(\ref{eq_1_svm}) produces a concave curve at $p>1$, where the condition
\begin{equation}
\frac{d^{2}\left(\log\eta(\widetilde{T})\right)}{d\left(\widetilde{T}_{g}/\widetilde{T}\right)^{2}}>0,
\end{equation}
is satisfied. At $p=1$, Eq.~(\ref{eq_1_svm}) produces a linear relationship between viscosity and inverse reduced temperature. Let us present experimental viscosity data as $(1/\alpha)\log[\eta(\widetilde{T})/\eta_{\infty}]$ vs. $\widetilde{T}_{g}/\widetilde{T}$ plot, using the values of the parameter $\alpha$ from Table~\ref{table_1}. Figure~\ref{fig_2}(b) shows that in this representation all experimental data are located inside a figure resembling a leaf of the tea tree. The tilted line delimits the ``fragile'' and ``strong'' glass-formers and this line is result of Eq.~(\ref{eq_1_svm}) at $p=1$. For example, the viscosity-temperature data for GeO$_2$ and As$_{2}$Se$_{3}$ are located along this boundary. SiO$_{2}$ and albite relating to the ``strong'' glass-formers are located above this line. Liquids relating to the ``fragile'' glass-formers are placed under this line [see Table~\ref{table_1}]. Thus, the parameter $p$ characterizes the degree of deviation of the temperature dependence of the viscosity from the Arrhenius law (red curve in Fig.~\ref{fig_2}(b)), which allow us to conclude that the parameter $p$ can be an analogue of the fragility index $m$.

Let us show that the proposed SV-model transforms to the Arrhenius law (\ref{eq_fa}). Taking into account (\ref{eq_Tscaling}), Eq.~(\ref{eq_1_svm}) can be written as
\begin{equation}\label{eq_vm_1}
\log\eta(T)=\log\eta_{\infty}+\alpha\left[2K_{1}(T_g,T_m)\left(\frac{T}{T_{g}}\right)+2K_{2}(T_g,T_m)\left(\frac{T}{T_{g}}\right)^{2}\right]^{-p}.
\end{equation} 
From Eqs.~(\ref{eq_m_index}) and (\ref{eq_vm_1}), we find the following expression for the fragility index:
\begin{equation}\label{eq_sc_m_index_1}
m=\alpha p(1+2K_{2}(T_g,T_m)).
\end{equation} 
Then Eq.~(\ref{eq_1_svm}) transforms to Eq.~(\ref{eq_fa}) at the condition
\begin{equation}\label{eq_vm_2}
\left[2K_{1}(T_g,T_m)\left(\frac{T}{T_{g}}\right)+2K_{2}(T_g,T_m)\left(\frac{T}{T_{g}}\right)^{2}\right]^{-p}=\frac{T_g}{T}.
\end{equation}
The condition (\ref{eq_vm_2}) is satisfied in two cases: (I) at $p=0.5$ and $T_g/T_m=0.707$ (or at $K_{2}(T_g,T_m)=0.5$ from Eq.~(\ref{eq_cond})); (II) at $p=1$ and $T_g/T_m=0.5$ (or at $K_{2}(T_g,T_m)=0$). Taking into account the value $\alpha\simeq17$ for an ideal ``strong'' glass-former, from Eq.~(\ref{eq_sc_m_index_1}) it follows that the both cases $p=0.5$ and $p=1$ correspond to the correct fragility index $m=17$. Hence, $p=0.5$ is the limit minimum value. For example, for SiO$_{2}$ we have $p=0.54$ and $T_g/T_m=0.725$, whereas these parameters take the values $p=0.51$ and $T_g/T_m= 0.779$ for albite [see Table~\ref{table_1}]. These values close to the case (I). The parameters $p$ and $T_g/T_m$ take the values $p=1.15$ and $T_g/T_m=0.594$ for GeO$_{2}$ that is close to the case (II). Thus, both cases are realizable.

\section{Discussion}

\begin{figure*}[ht]
	\centering
	\includegraphics[width=13cm]{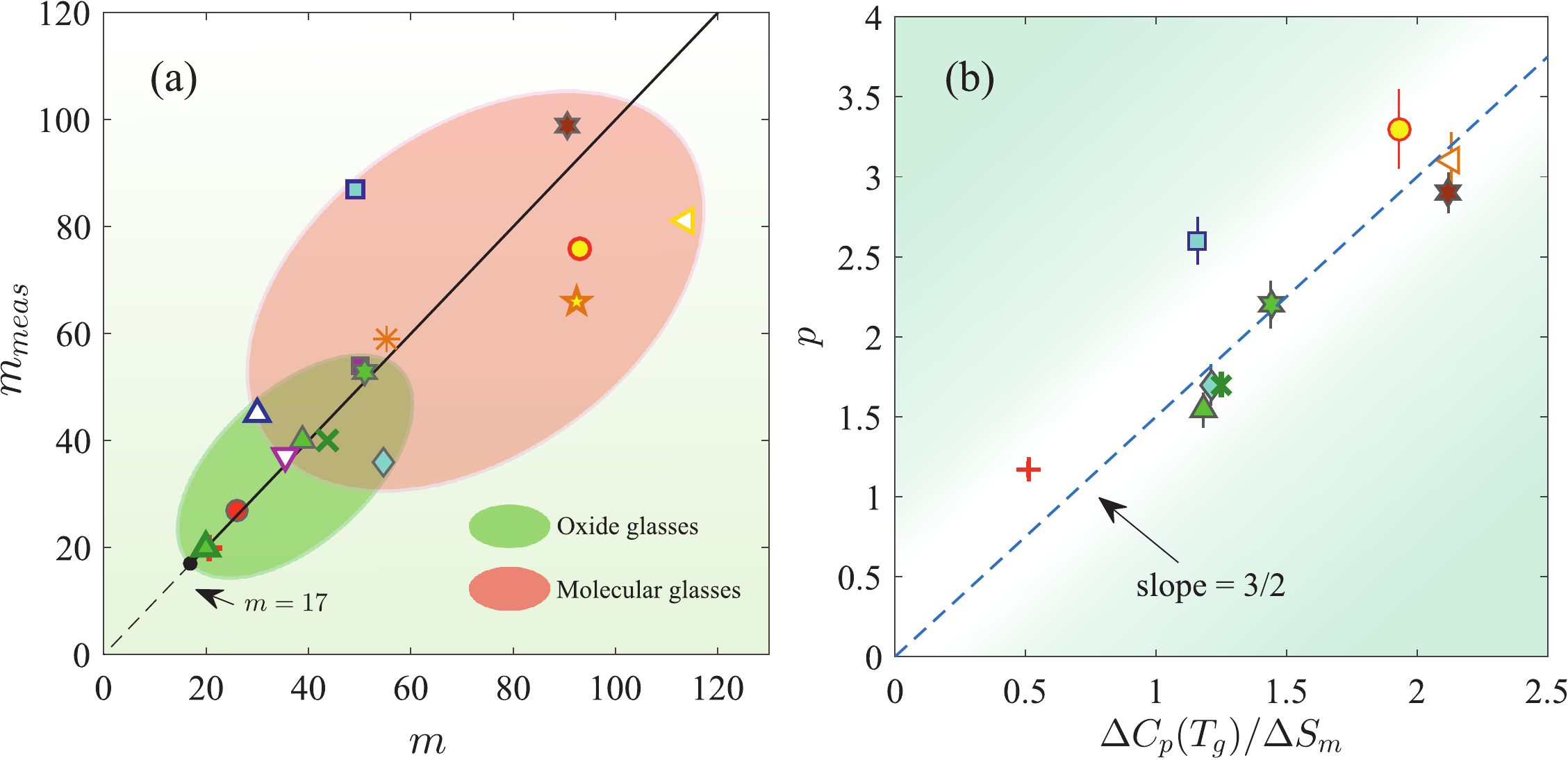}
	\caption{(a) Correspondence between fragility index $m$ calculated by Eq.~(\ref{eq_sc_m_index_1}) and experimentally measured fragility index $m_{meas}$ from Refs.~\cite{Wang_Angell_Richert_2006,Angell_Plazek_1993,Senkov_Miracle_2009} [see Table~\ref{table_1}]. (b) $p$ versus $\Delta C_{p}(T_g)/\Delta S_{m}$ plot, where values of these parameters are taken from Table~\ref{table_1}.}\label{fig_3}
\end{figure*}

The fragility index $m$ calculated through Eq.~(\ref{eq_sc_m_index_1}) was compared with the experimentally measured fragility index $m_{meas}$ available for some glass formers. The values of the parameter $m_{meas}$ are taken from Refs.~\cite{Wang_Angell_Richert_2006,Angell_Plazek_1993,Senkov_Miracle_2009} and given in Table~\ref{table_1}. Fig.~\ref{fig_3}(a) shows a good agreement between the parameters $m$ and $m_{meas}$ in the case of oxide glasses. In the case of molecular glasses such as TNB, salol and o-terphenyl, there is a divergence between the values of these parameters. The main reason for this divergence is the difference in the conditions for the experimental determination of the viscosity and thermodynamic parameters of these fragile systems, on the basis of which the fragility indices $m$ and $m_{meas}$ are calculated. Thus, Eq.~(\ref{eq_sc_m_index_1}) leads to the correct values of the fragility index $m$, which in turn shows the correctness of the viscosity model presented in the form of Eq.~(\ref{eq_vm_1}).

Based on the available experimental data from~\cite{Wang_Angell_Richert_2006}, we have found a linear relationship between the parameter $p$ and the thermodynamic ratio $\Delta C_{p}(T_g)/\Delta S_{m}$, where $\Delta C_{p}(T_{g})$ is the heat capacity differences of supercooled liquids and $\Delta S_{m}$ is the entropy of fusion. Fig.~\ref{fig_3}(b) shows that values of the parameter $p$ increase linearly with increasing $\Delta C_{p}(T_g)/\Delta S_{m}$. The simplest estimate of this relationship can be performed using the linear function with the slope $3/2$:
\begin{equation}\label{eq_new_m}
p=\frac{3}{2}\frac{\Delta C_{p}(T_g)}{\Delta S_{m}}.
\end{equation}
As seen in Fig.~\ref{fig_3}(b), some points fall outside the linear law on the $p$ vs. $\Delta C_{p}(T_g)/\Delta S_{m}$ plot. This may be due to measurement mistakes that arise when evaluating the values of these parameters. At this stage of our considerations, let us neglect an impact of these points in the correlation between $p$ vs. $\Delta C_{p}(T_g)/\Delta S_{m}$. Then, taking into account Eq.~(\ref{eq_new_m}), the expression for the fragility index (\ref{eq_sc_m_index_1}) takes the form
\begin{equation}\label{eq_new_m1}
m\simeq f\frac{\Delta C_{p}(T_g)}{\Delta S_{m}}, 
\end{equation}
where
\begin{equation}\label{eq_new_m2}
f=\frac{3}{2}\alpha\left[1+2K_{2}(T_g,T_m)\right].
\end{equation}
It is noteworthy that the value of the parameter $f$ averaged over all considered systems is $<f>\approx38.5$. This value is close to the value $<f>=40$ from Ref.~\cite{Wang_Angell_Richert_2006}. According to Ref.~\cite {Wang_Angell_Richert_2006}, the ratio $\Delta C_{p}(T_g)/\Delta S_{m}$ is the so-called thermodynamic fragility index, which determines the slope in $\Delta S (T)/\Delta S_{m} $ vs. $T/T_{m}$ plot at $T=T_{g}$, where $\Delta S$ is the glass-to-liquid entropy difference. Then $p$ is the parameter connecting the kinetic fragility index $m$ with the thermodynamic fragility index $\Delta C_{p}(T_g)/\Delta S_{m}$.    

\begin{figure*}[ht]
	\centering
	\includegraphics[width=15.0cm]{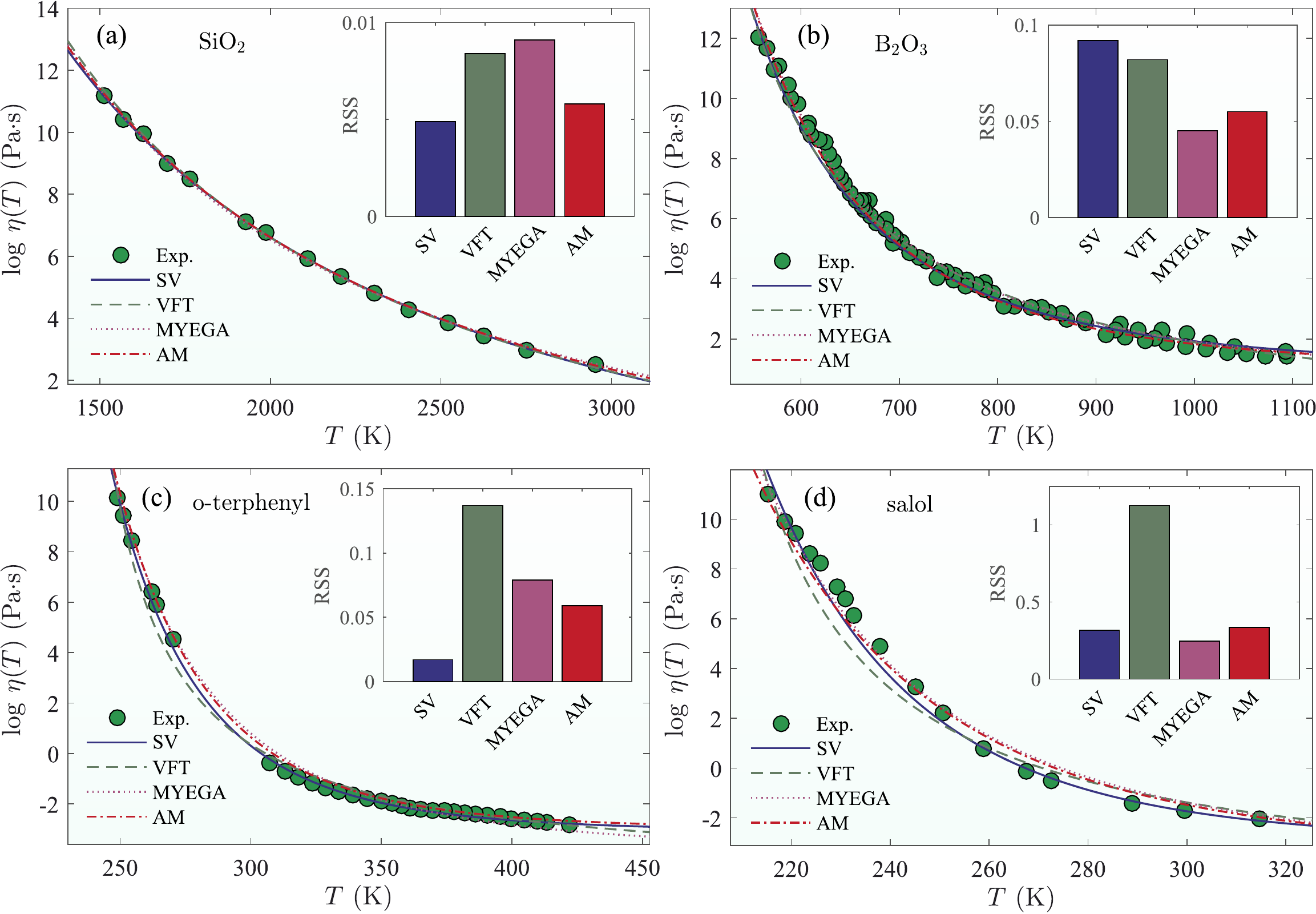}
	\caption{Comparison of SV, VFT, MYEGA and AM viscosity models as applied to (a) SiO$_{2}$, (b) B$_{2}$O$_{3}$, (c) o-terphenyl and (d) salol. Insets: mean residual sum of squares (RSS) calculated by Eq.~(\ref{eq_rss}).}\label{fig_4}
\end{figure*}

In Fig.~\ref{fig_4}, we compare the proposed scaled viscosity model (\ref{eq_vm_1}) with the VFT, AM and MYEGA viscosity models. The discrepancy between the experimental data and the results of these viscosity models was calculated as the mean residual sum of squares (RSS):
\begin{equation}\label{eq_rss}
RSS=\frac{1}{n}\sum_{i=1}^{n}\left[\eta_{i}^{(Exp)}(T)-\eta_{i}^{(Model)}(T)\right]^{2}.
\end{equation}
Here, $n$ is the number of experimentally measured points in $\eta$ vs. $T$ plot. The lower the RSS value, the more accurately the model reproduces the experimental data. As an example, four different glass-formers (SiO$_{2}$, B$_{2}$O$_{3}$, o-terphenyl and salol) are considered, which experimental data are available for a wide temperature range. These experimental data were reproduced by the viscosity models VFT [Eq.~(\ref{eq_vft_model})], MYEGA [Eq.~(\ref{eq_myega_model})], AM [Eq.~(\ref{eq_am_model})], and the proposed SV-model [Eq.~(\ref{eq_vm_1})].
As seen in Fig.~\ref{fig_4}(a), all the viscosity models correctly reproduce the experimental data for SiO$_{2}$. Compared to other viscosity models, the SV-model gives better agreement with experiment, as evidenced by the lowest RSS value [see inset on Fig.~\ref{fig_4}(a)]. Fig.~\ref{fig_4}(b) shows that the VFT and SV-models give close RSS values for B$_{2}$O$_{3}$ and these values are larger than the RSS of the MYEGA and AM viscosity models. The reason of large RSS is the discrepancy between the experimental data and the viscosity models near the glass transition temperature. In the case of o-terphenyl and salol, the SV-model is more accurate than the VFT-model. From the insets to Figs.~\ref{fig_4}(c) and~\ref{fig_4}(d), it can be seen that the RSS value for the VFT-model is the highest among the other viscosity models due to poor approximation of experimental data. This indicates that the value of the fragility index $m$ calculated for molecular liquids based on the parameters of the VFT equation can vary greatly. The presence of three adjustable parameters in the VFT equation defines a wide range of possible values for $m$. Notable is the high accuracy of the two-parameter SV-model, which allows one to significantly narrow the range of possible values of the fragility index for a system. This explains the discrepancy between $m$ and $m^{(VFT)}$ calculated using SV and VFT equations.

\section{Conclusions}

In the present work, a description of viscosity-temperature data for the different glass-forming liquids is performed using the proposed scaled viscosity model. It is shown that the equation of SV-model is able to reproduce experimental viscosity data using only two fitting parameters. The comparison with other viscosity models including the VFT equation shows the high accuracy of the proposed model, which is confirmed by relatively low mean residual sum of squares. This accuracy leads to values of the fragility index that differ from the values obtained in the framework of the VFT-model. For example, significant changes in the values of the fragility index are observed in the case of o-terphenyl, butylbenzene and Ti$_{40}$Zr$_{10}$Cu$_{30}$Pd$_{20}$ alloy: the estimated fragility index is more than $1.3$ times higher than the data on the VFT-model. Moreover, the scaling description using a reduced uniform temperature scale made it possible to clearly distinguish between the ``strong'' and ``fragile'' glass-formers. The obtained results make it possible to expand the idea of a unified description of the temperature dependent physical characteristics of the crystallization kinetics using unified scaled relations.

\section*{Conflicts of interest}
\noindent The authors declare that they have no conflict of interest.

\section*{Acknowledgement}
This study is supported by the Russian Science Foundation (project No. 19-12-00022).

\bibliographystyle{unsrt}

\end{document}